%% file: main.tex
\documentclass[10pt,conference]{IEEEtran}

\input{notations}

\usepackage[disable]{todonotes}

\begin{document}

\title{COOK Access Control on an embedded Volta \gpu} 

\author{
\IEEEauthorblockN{Benjamin Lesage}
\IEEEauthorblockA{\textit{ONERA}}
Toulouse, France \\
benjamin.lesage@onera.fr
\and
\IEEEauthorblockN{Fr\'ed\'eric Boniol}
\IEEEauthorblockA{\textit{ONERA}}
Toulouse, France \\
frederic.boniol@onera.fr
\and
\IEEEauthorblockN{Claire Pagetti}
\IEEEauthorblockA{\textit{ONERA}}
Toulouse, France \\
claire.pagetti@onera.fr
}

\maketitle

\begin{abstract}
 The last decade has seen the emergence of a new generation of \multicore 
in response to advances in machine learning, and in particular Deep Neural Network (DNN) training and inference tasks. 
These platforms, like the \jetsoncomplet, embed several cores and accelerators in a SWaP-efficient (Size Weight and Power) package with a limited set of resources.
However, concurrent applications tend to interfere on shared resources, resulting in high execution time variability for applications compared to their behaviour in isolation.

\emph{Access control} techniques aim to selectively restrict the flow of operations executed by a resource.
To reduce the impact of interference on the \jetson Volta \gpu, we specify and implement an access control technique to ensure each \gpu operation executes in isolation to reduce its timing variability.
We implement the controller using three different strategies and assess their complexity and impact on the application performance.
Our evaluation shows the benefits of adding the access control: 
its transparency to applications, reduced timing variability, isolation between \gpu operations, and small code complexity. 
However, the strategies may cause some potential slowdowns for applications even in isolation but which are reasonable.
\end{abstract}


\input{sec_introduction}
\input{sec_platform}

\input{sec_context}
\input{sec_specification}
\input{sec_resource_manager}
\input{sec_methodology}
\input{sec_evaluation}
\input{sec_conclusion}

\section*{Acknowledgement}
\label{sec:acknowledgment}

The work presented in this paper is part of the PHYLOG 2 project supported by the Directorate General of Civil Aviation (DGAC). It is funded by the French government through the France Relance program, based on the funding from the European Union through the NextGenerationEU program.



\bibliographystyle{IEEEtran}
\bibliography{bib}


\end{document}

%% file: notations.tex
\usepackage[utf8]{inputenc} 
\usepackage{hyperref}
\usepackage{xspace}
\usepackage{enumitem}
\usepackage{url}
\usepackage{tikz}                
\usepackage{cite}                
\usepackage{subcaption}      
\usepackage{multirow}         
\usepackage{algpseudocode} 
\usepackage{algorithm}        

\usetikzlibrary{shapes,shapes.geometric,backgrounds}
\usetikzlibrary{calc} 

\newcommand{\multicore}{multi-core\xspace}

\newcommand{\gpu}{GPU\xspace}
\newcommand{\gpus}{GPUs\xspace}

\newcommand{\jetsoncomplet}{{\sc Jetson AGX Xavier}\xspace}
\newcommand{\jetson}{{\sc Jetson}\xspace}

\newcommand{\carmel}{{\sc Carmel}\xspace}

\newcommand{\nvidia}{{\sc Nvidia}\xspace}
\newcommand{\otawa}{{\sc Otawa}\xspace}
\newcommand{\ait}{{\sc aiT}\xspace}

\newcommand{\mastercode}{host code\xspace}

\newtheorem{requirement}{Aspect}

\newcommand{\benchmmult}{\texttt{cuda\_mmult}\xspace}
\newcommand{\benchdna}{\texttt{onnx\_dna}\xspace}
\newcommand{\benchcfg}[3]{\texttt{#1-#2-#3}}
\newcommand{\slowdown}[1]{$#1\times$}

%% file: sec_introduction.tex

\section{Introduction}
\label{sec_introduction}

%
%

The aeronautic domain faces two constraints:
first, it is subject to certification
and second, it relies massively on Commercial Off-The-Shelf~(COTS) processors.
Thus, to embed any \multicore processor -- with or without accelerators --,
it is mandatory to provide assurance of 
 \emph{time predictability}.
Time predictability encompasses the capability to compute a safe and tight upper bound on the number of cycles, the \emph{execution time}, required to execute an application in the worst case \cite{Wilhelm2008,Wilhelm2012}.
Time predictability implies mastering the hardware platform mechanisms and their impact on execution time variability.
Indeed, high variability may endanger the overall aircraft/system safety an thus should thus be avoided as much as possible.
This is notably the case in the presence of interference, where applications compete for shared resources resulting in additional variability over their execution time in isolation.

\paragraph*{\textbf{Context}}
COTS platforms, \multicore- or accelerator-based, tend to lack a clear documentation which hinders the analysis efforts to understand and bound the effect of interference. 
They further tend to neglect predictable hardware and software mechanisms, in favour of delivering higher throughput.
While they may exhibit little variability for applications in isolation, interference and uncertainty tend to be exacerbated when applications execute in parallel.
\emph{Mitigation} techniques aim to reduce or bound interference occurring on a platform, and as a result the related timing variability. 
\todo{Add general ref for access control}
\emph{Access control} techniques in particular operate by regulating the flow of operations to specific resources, to reduce pressure on them and provide finer-grain control.
Many papers tackle this problem for \multicore, but far less focus on accelerators. 

\begin{figure*}[hbt]
\center
    \includegraphics[width=0.28\linewidth]{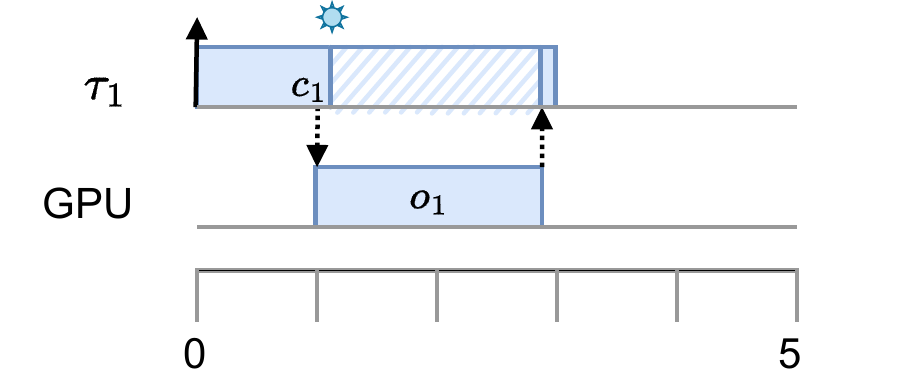}
    \includegraphics[width=0.28\linewidth]{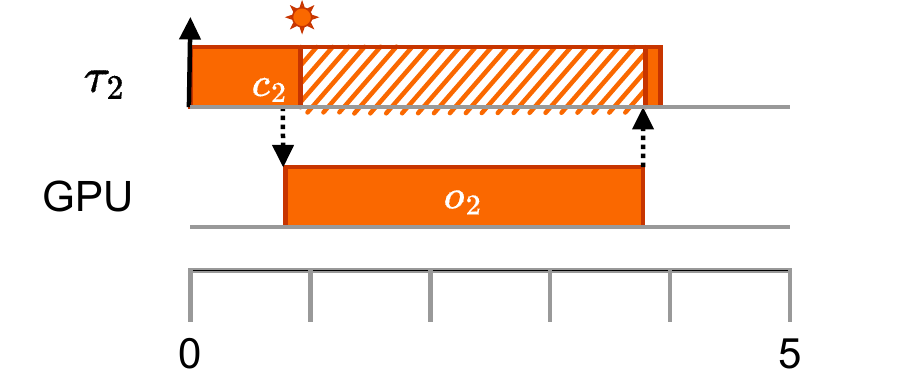}
    \includegraphics[width=0.28\linewidth]{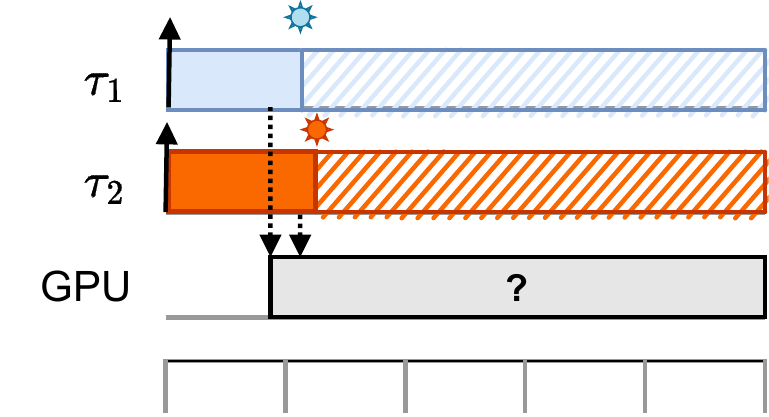}
    \caption{Applications 1 and 2 running a kernel in \emph{isolation} (resp. left and centre), or in \emph{parallel} on the \gpu (right).}
    \label{fig:diag_burst}
  \end{figure*}

Consider an application, as shown in Figure \ref{fig:diag_burst} (left), composed  of
\emph{\mastercode} that executes on a CPU core and offloads some
\emph{operations} onto the \gpu. 
Application $\tau_1$ offloads a single operation, and waits for the completion of the operation (star-shaped synchronisation point).
Figure \ref{fig:diag_burst} (centre) shows another application also offloading some \gpu operation.
When two applications run in parallel, \mastercode execution on the CPU is handled by the Operating System (OS) scheduler.
At some point, \gpu operations from both applications are offloaded to the \gpu.
Figure \ref{fig:diag_burst} (right) highlights this behaviour, as well as our
questions:
1) What is the internal behaviour of the operations and how do they interfere?
2) What is the impact for an application on its execution time to run in parallel, over isolation?
3) How to account for, or mitigate said impact if not acceptable?

\paragraph*{\textbf{Contributions}}

Our objective is to mitigate the timing variability on \gpu operations generated by concurrent applications. 
We propose a temporal access control technique to isolate the execution of said operations on the \gpu, reducing the impact of interference on the operations' execution time variability.
The method generates software hooks~\cite{api_hooking} from simple templates to modify the behaviour of existing \gpu routines.
The templates can thus easily be adapted to new or updated routines.
Using hooks, the access controller is transparent to applications; it requires no modification of applications and can support a wide variety of runtime environments.
Our target platform, the \jetsoncomplet~\cite{NVIDIA_Xavier_datasheet}, is introduced in Section~\ref{sec_platform} to outline its behaviour w.r.t. the \gpu, source of interference, and mitigation means. 
Section~\ref{sec_context} describes related work in the context of the \jetson.
Based on these observations, Section~\ref{sec_specification} highlights constraints and desirable features for our access controller.
Section~\ref{sec_resource_manager} then introduces our access control technique, 
and compliant implementations. 
The evaluation in Sections~\ref{sec_methodology} and~\ref{sec_evaluation} assess our claim that the proposed approach satisfies its objectives: transparency for the application, isolation of \gpu operations, and mitigation of \gpu operations timing variability.

%% file: sec_platform.tex

\section{Jetson AGX Xavier platform}
\label{sec_platform}

This section presents the Volta GPU on the \jetsoncomplet with a focus on its execution model, to identify notable resources, sources of interference, and possible means of mitigation.


\subsection{System model}

\begin{figure*}[hbt]
	\centering
	\includegraphics[width=0.68\linewidth]{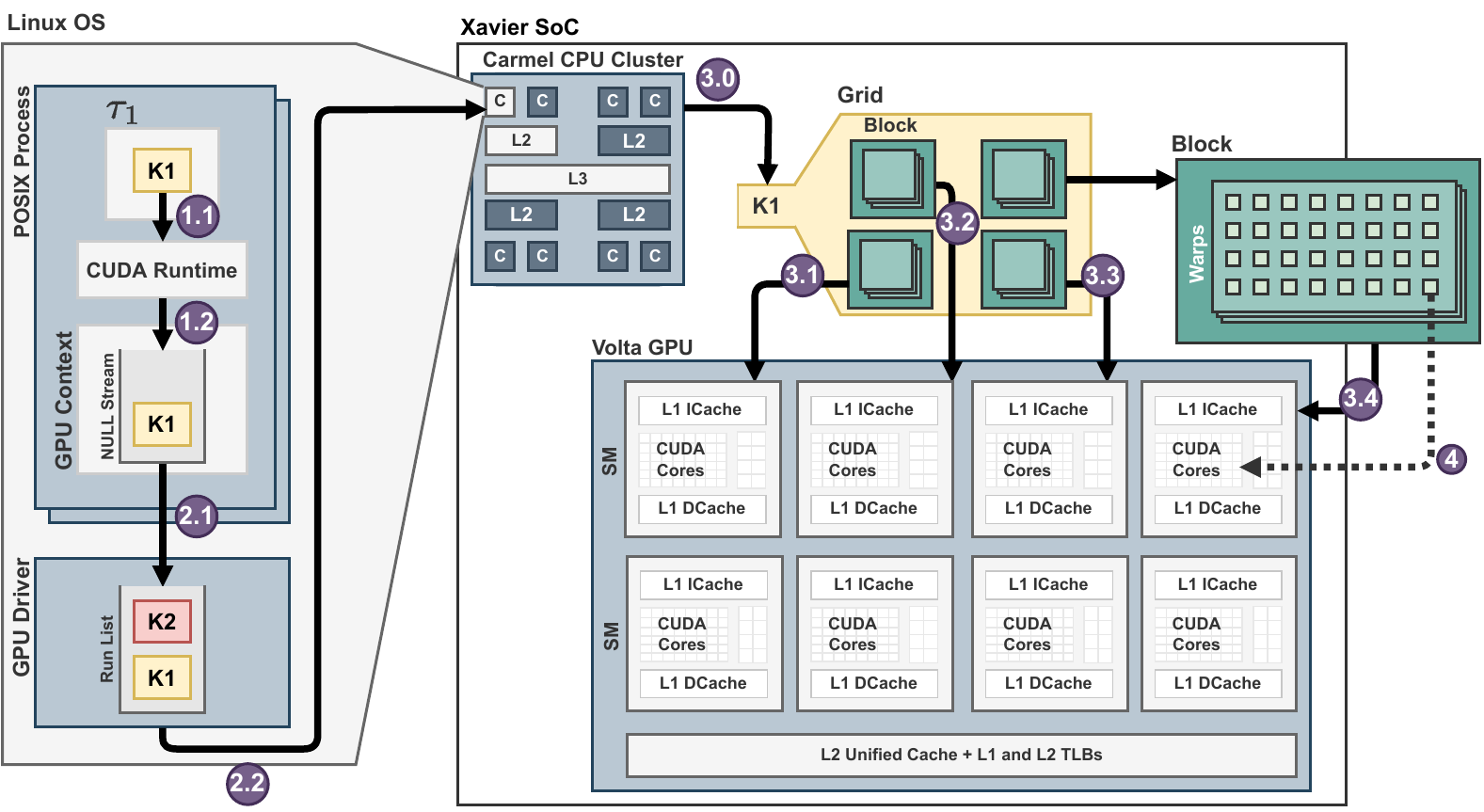}
	\caption{Overview of the Volta \gpu execution model 
                 \label{fig-cuda_lifecycle}}
  \hfill
\end{figure*}

We consider a simple system where one or more applications $\tau_i$ execute on the \jetson.
It features the Xavier System-on-Chip (SoC) outlined in Figure~\ref{fig-cuda_lifecycle}. The Xavier SoC embeds a Volta GPU, an 8-core \carmel CPU complex, and dedicated accelerators for video and audio processing applications (omitted from the Figure). 
The CPU complex and \gpu access the main memory through a shared memory controller fabric.
In this work, we assume that only one of the \carmel cores per application is in use and that
the other accelerators are not.
Each application furthermore uses its own separate \gpu context and stream to enqueue \gpu operations (left of Figure~\ref{fig-cuda_lifecycle}).



\begin{figure}[hbt]
	\centering
    \includegraphics[width=0.58\linewidth]{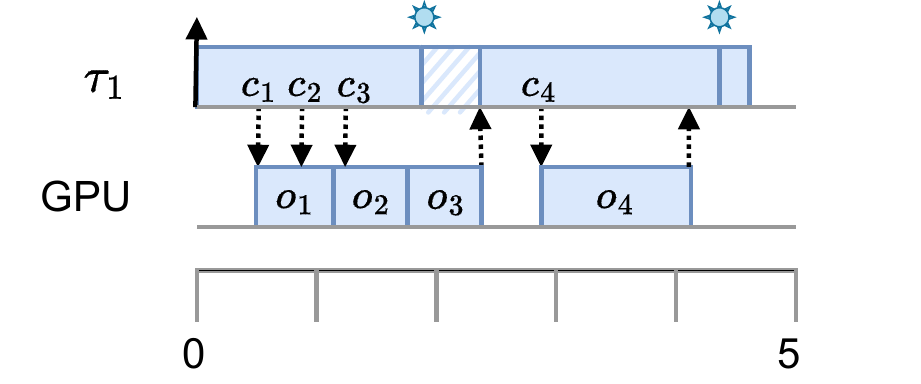}
		\includegraphics[width=.40\linewidth]{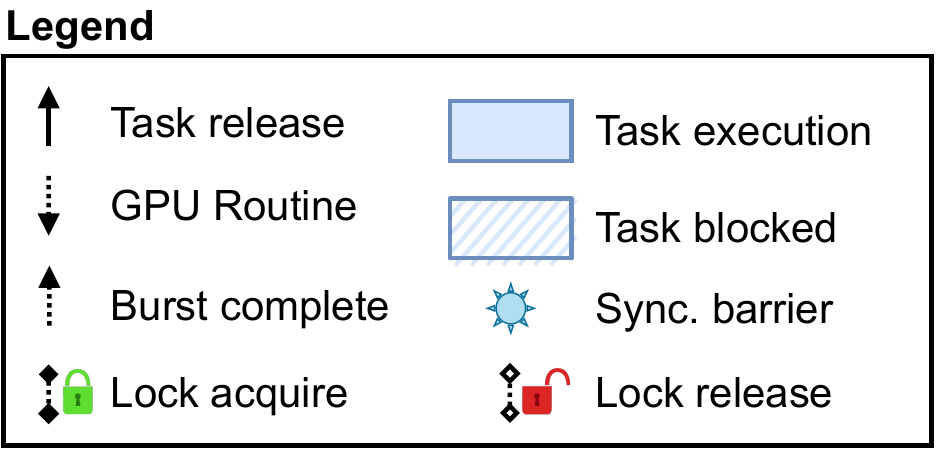}
    \caption{Single application with two kernel bursts 
    \label{fig-burst-application}}
\end{figure}

  An application $\tau_i$, as illustrated in the Figure~\ref{fig-burst-application} chronogram, is composed of two parts:
  \begin{itemize}[noitemsep,topsep=0pt]
  \item a \emph{\mastercode} that executes on the \carmel cores and offloads work on the \gpu itself through \emph{\gpu routines};
  	\begin{itemize}[noitemsep,topsep=0pt]
  		\item 
  		  A \emph{\gpu routine} is an asynchronous call $c_i$ on the host (dashed downward arrow) to execute one          
  		  corresponding \emph{\gpu operation} $o_i$; 
  		\item 
  		  The \mastercode may perform a number of host-side operations 
  		  in-between calls to \gpu routines;
  		\item 
  		  A \emph{synchronisation barrier} (star symbol) is a synchronous \emph{\gpu routine} 
  		  awaiting for the completion of all prior \gpu operations. The barrier may thus block 
  		  the application (hashed blocks).
  	\end{itemize}
    \item a sequence of \emph{\gpu operations}, $o_i$, organised in \emph{bursts}:
      \begin{itemize}[noitemsep,topsep=0pt]
      \item 
        a \emph{burst} is a sequence of \emph{\gpu operations} 
        launched as an ordered group of \gpu routines by the \mastercode;
      \item
        a \emph{\gpu operation} is either a \emph{kernel}, or a \emph{copy}
        operation that runs on the \gpu;
        \begin{itemize}[noitemsep,topsep=0pt]
        \item 
          a \emph{copy} operation moves data between the host and \gpu memory space;
        \item
          a \emph{kernel} operation is a function executed on the \gpu following a \emph{grid}, 
          that is groups of threads, where a thread is the basic \gpu execution unit running the kernel in parallel;
        \end{itemize}
      \item the \emph{bursts} are ordered in a sequence:
        the \mastercode sends a burst of operations,
        then waits for the results from the \gpu using a \emph{synchronisation barrier}.
        Only after the completion of the burst (dashed upward arrow), the application then sends the next burst of 
        operations waiting for its results, and so on until all the bursts have been offloaded.
      \end{itemize}
  \end{itemize}

The designer has many ways to architect their applications.
Indeed, they need to choose all the application parameters:
number, type, and order of \gpu operations, number of bursts, size of bursts, and position of synchronisation barriers.
Figure~\ref{fig-burst-application} shows such a trade-off for an application running in isolation composed of two bursts, of respectively three ($o_1$, $o_2$, $o_3$) and one ($o_4$) operation(s).
In the first burst, the CPU has to wait for the completion of its \gpu operations. The second burst of \gpu operations completes before the synchronisation barrier, and the CPU can immediately cross said barrier.


\subsection{Volta Execution Model}
\label{sec_execution_model}

A user can define and execute kernels and other \gpu operations through the CUDA Runtime or rely on high-level libraries, on top of the Runtime, such as cuDNN.
The operations transit through the CUDA Runtime to the driver (left in Figure~\ref{fig-cuda_lifecycle}) which interacts with the \gpu to setup the required structures. 
Most calls to the CUDA Runtime from the \mastercode are asynchronous, allowing applications to submit a burst of \gpu operations until synchronisation is required, e.g. to retrieve some results from the \gpu.
We focus in the following on the definition and execution of computation kernels.


A kernel call is shaped by the computation \emph{grid} definition, i.e. the number of \emph{thread blocks} and their shape. All thread blocks in a call are equally shaped~\cite{NvidiaCudaThreads}. 
The size of the kernel, the total number of threads invoked on a call, is thus the product of the number of threads per block and the number of blocks. 
Each thread is given a unique identifier to address different data segments. It is composed of its block index in the grid, and its thread index in the block. Both are accessible during kernel definition through the CUDA C extensions.

A kernel call is placed in a \emph{stream}, a (possibly) user-defined queue of GPU operations (Steps 1.1 and 1.2 in Figure~\ref{fig-cuda_lifecycle}). This provides some guarantees on the ordering of related operations with regards to their execution on the GPU. The call travels through the software stack (Steps 2.1 and 2.2) through multiple queues. While the First-In First-Out~(FIFO) ordering of operations in a stream is maintained, a kernel might be interleaved with kernels from other streams and run concurrently with them. The block scheduler (Steps 3.0--3.4) dispatches each block of the kernel to a Streaming Multiprocessor (SM), depending on the block's resource requirements and the SM occupancy. The block will remain on its allocated SM until its threads complete (unless a rescheduling occurs on preemption).

The \jetsoncomplet Volta \gpu is composed of 8 SMs, depicted in Figure~\ref{fig-cuda_lifecycle}, each with its own private L1 instruction and data caches (respectively ICache and DCache). All SMs share a unified L2 cache and 2 levels of TLB. Each SM is further divided into 4 processing blocks (SMP). The register file on an SMP, or L0 data cache, holds the context of multiple threads. Each SM is itself composed of 64 CUDA cores and 8 Tensor cores, split evenly across the SMP. Tensor cores are a class of Deep Learning Accelerators supporting multiply-add operations on matrices, as instructed by the user (or a library). Restrictions on the number of registers in each SM imply that not all threads in a kernel call can execute concurrently. Blocks can hold a maximum of 1024 threads, and at most 32 blocks can reside at once on a single SM.

SMs follow the Single Instruction Multiple Data (SIMD) execution model: one instruction is executed across multiple threads at the same time, possibly addressing different data\footnote{Note that the Volta GPU architecture introduces thread divergence, where threads may have different program counters. The scope and impact of thread divergence on the SIMD model is unclear.}. Threads are scheduled and executed in groups of 32, called \emph{warps}, such that each thread belongs to a single warp across its lifetime.  Warp size and the allocation of threads in warps are not user-controlled parameters but platform specific. The warp scheduler on each SMP can schedule up to one warp every cycle. Instructions from two distinct warps may coexist on the same core provided they rely on different functional units, e.g. a long running load due to a cache miss and an integer addition. The combined register files on each SM can accommodate up to 64 warps to ensure the warp schedulers can maximise core occupancy.

%% file: sec_context.tex

\section{Related work}
\label{sec_context}

There has been considerable effort to characterise the behaviour of GPU accelerators, in particular work on \nvidia \gpus \cite{charac:Volta, charac:pastis, charac:Nathan,charac:cuda_sched} and the assorted CUDA software stack \cite{charac:pitfalls, charac:Nathan, charac:Tanya, charac:cuda_sched}.
\cite{charac:Tanya, charac:Nathan} do identify high-level scheduling rules for kernels in the CUDA software stack, down to the dispatching of their constituent blocks on the \gpu resources. These rules notably highlight that software resources might induce interference between concurrent kernels.
These contributions highlight the need for mitigation techniques to cope with interference on \gpu.

\subsection{Related Work on Access Control for GPU accelerators}

Various resource managers have been proposed in the context of multi- and many-core architectures to build robust, reconfigurable platforms, e.g. in the context of the SCARLETT~\cite{rm:SCARLETT}, ACTORS~\cite{actors}, DREAMS~\cite{rm:DREAMS}, or SECREDAS~\cite{rm:SECREDAS} projects. 
They allow for a unified view of the resources available across the whole system.
All these methods share the same requirement for robust access control methods: to understand the available resources' capacity to serve queries without causing interference, to allocate a portion of said capacity, and to possibly reclaim said capacity at runtime. 

The Volta GPU features a wide range of resources which might act as interference sources between between concurrent operations. 
Unfortunately, the Volta GPU architecture offers no granularity of isolation between kernels. 
\nvidia proposes Multi-Instance \gpu (MIG)~\cite{NVIDIA_mig} to allocate 
hardware resources to a kernel. 
However, MIG is supported by neither the Volta \gpu nor the \jetson platforms. 
We thus consider software access control techniques, in the absence of explicit hardware support. They operate on \emph{who} does execute the kernel, and \emph{when} to execute kernels. They may rely, or not, on the application \emph{cooperation} to the access control by calling specific primitives or restricting its demand on the resource capacity. 

The approach in \cite{charac:Nathan} did consider the use of access control techniques on AMD platforms, comparing locking and dedicated \gpu support. 
Although AMD \gpus do explicitly support mapping kernels to specific cores, 
such facilities are not officially supported on the \jetson product line.
\cite{rm:prt_nvidia} did recently highlight through patent analysis and reverse-engineering that \nvidia \gpus have had the capability for a decade. 
The authors do offer an API to exploit it on \jetson platforms and they achieve compute resource partitioning on \jetson platforms. 
The approach does rely on cooperative applications, and it is limited to CUDA versions which have been analysed.

Temporal access control techniques 
restrict the kernel(s) 
which can run on the GPU at any given time (\emph{when}). 
\emph{Locking}~\cite{rm:TimeWall, rm:bwlock, rm:gpusync} relies on applications acquiring a shared lock before accessing the resource to guarantee mutual exclusion, either at the \gpu-level~\cite{rm:TimeWall} or between the \gpu and CPU~\cite{rm:bwlock}. 
Similarly, a server \cite{rm:Server, rm:GPU_preemption, rm:rgem} acts as a concentrator or proxy for applications' requests, scheduling kernels and redirecting them to the resource. 
RGEM~\cite{rm:rgem} in particular replaces the \gpu libraries with a custom implementation to schedule \gpu operations in isolation.
Locks and servers intervene on the CPU-side, before operations are dispatched to the \gpu. 


Persistent Thread Blocks (PTB) \cite{rm:PTB, rm:PTB_capacity, rm:FGPU} provide a spatial access control method to allocate compute resources~(SM) on the GPU. PTB act as runners for a kernel (\emph{who}).
A PTB executes the kernel's blocks, fetched from a work queue, as its own. 
It will persist, on a specific SM, repeating the process until completion or otherwise instructed.
Reducing the number of PTB for a kernel thus reduces the kernel's demand on the GPU capacity. 
PTB however require a cooperative application, modifying its kernels to wrap their definition and execution into the PTB execution loop. 
Such modifications can be performed during kernel definition or automatically, as source-to-source transformations.

We focus in the following on locking-based approaches. 
They have no impact on the \gpu software stack.
Furthermore, each application remains in charge of its own accesses to the \gpu.
In comparison, a server acts as a unique concentrator for \gpu operations.
Consider as an example the \nvidia Multi-Process Service (MPS) server~\cite{nvidia:mps}. Its error containment is such that a fatal exception in a kernel may propagate to others sharing the same GPU. 
We also leverage the principle of library hooking to offer a platform to deploy different mitigation strategies, even in presence on non-cooperative applications.

\subsection{Related Work on CUDA Hooking}

A hook~\cite{api_hooking} is a function which (often dynamically) replaces an existing one in an application or library. Hooking may be used in a wide range of contexts to extend applications when recompilation is not an option, notably for providing debugging, instrumentation, or analysis capabilities. \nvidia relies on hooking for a number of its tool. \texttt{nsys} does inject replacement CUDA libraries into an instrumented application to trace their execution and collect relevant statistics. However, \texttt{nsys} does not allow for users to define their own hooks. This was supported by NVBit~\cite{charac:nvbit}: user-defined tools are loaded as dynamic libraries to intercept calls to the CUDA stack. However the tool is no longer maintained.

BWLOCK \cite{rm:bwlock} replaces CUDA API calls to automatically acquire/release the bandwidth lock upon kernel execution. RGEM~\cite{rm:rgem} entirely replaces the CUDA API implementation with its own. Similarly, numerous works focused on \gpu virtualisation~\cite{rm:gvirtus, rm:rcuda, rm:dscuda} provide replacements for the CUDA API to defer the execution of \gpu kernels to remote hosts. An application can thus transparently execute kernels from a host without a \gpu. However, CUDA hooks tend to be ad-hoc replacements, for only a limited subset of the CUDA API. Applications are thus limited to the supported methods. Furthermore, some CUDA libraries circumvent the hook injection methods used by these approaches.

GPUSync~\cite{rm:gpusync} does address similar scheduling concerns as our approach, considering \gpu and copy engines allocation, worker preemptions, and closed-source driver threads.
The approach further benefits from its integration into the platform kernel, currently unsupported by the \jetson product line.
Information on the hooks implemented by GPUSync is sparse, and it is unclear if it could support a runtimes such as the ONNX one which circumvent some library loading mechanisms.
Our contribution supports (without aiming for) the definition of a scheduler, e.g. generating hooks for approaches like GPUSync.

%% file: sec_specification.tex

\section{Access control on the \jetson}
\label{sec_specification}

The behaviour and control offered by the \jetsoncomplet platform (Section~\ref{sec_platform})  restricts the scope of applicable techniques,
to mitigate the effect of interference on the execution time of the \gpu operations of an application. 
We also need to  consider lessons learned from existing work (Section~\ref{sec_context}). 


\subsection{Constraints}

Considering the kernel execution model in Figure~\ref{fig-cuda_lifecycle} highlights a number of shared resources on the Volta \gpu,
both at the software (left) level, and hardware (right) one.
We need to consider potential interference sources, support for any granularity of access control, and understand the level of control available on the platform. \todo{This sentence could go}

Cores (their registers and functional units) are split evenly between SMs, each with its own private L1 caches.
Any method allocating resources to applications at the SM level would still suffer interference on the shared L2 and TLB caches.
Due to a lack of documentation, we did not represent any communication medium inside the \gpu.
However, prior work~\cite{charac:cuda_sched} did highlight that concurrent accesses to the same resource may suffer interference due to a shared communication medium.
The software stack suffers from a similar lack of clarity in spite of recent reverse-engineering efforts~\cite{charac:Tanya}.
All layers of the platform, from the libraries to the \gpu, may contribute to some extent to scheduling the \gpu operations between applications.
A \emph{stream} does provide guarantees on the ordering of its own \gpu operations, but few on the ordering or isolation of operations between streams in the same or separate \gpu contexts.
Separate OS processes do default to separate \gpu contexts, thus providing some isolation. 
But the software stack still funnels \gpu operations into a limited number of shared queues, within and between \gpu contexts, and then at the driver level.


The \gpu hardware offers only limited visibility and control on the resources used by or allocated to a running \gpu operation, and no granularity of isolation between operations.
Similarly at the application level, once a \gpu operation has been inserted into a stream the application loses control over its execution or schedule.
In the absence of a formal platform model or analysis tool, such as \otawa~\cite{BallabrigaCRS10}, PasTiS~\cite{charac:pastis} or \ait~\cite{wcet:ait}, each \gpu operation is considered as a non-preemptable black box, and we rely on end-to-end execution timing measurements.
We further focus on software-level techniques, in the absence of hardware support.
We discuss target features in the following section.




\subsection{Specification}
\label{sec_requirements}


We consider a number of desirable aspects for the proposed access control technique. As discussed earlier, those consider related work, their benefits and drawbacks.

\begin{requirement}[Transparency]
\label{requirement_transparency}
The method should be transparent to the application. It should require no modification to the application source code or model, nor cooperation with a specific API, nor modification to its runtime if any. The aspect notably ensures the method can be applied to a wide range of applications, or models without restriction on their runtime or additional maintenance.
\end{requirement}

\begin{requirement}[Resiliency]
\label{requirement_simplicity}
The method should require no modification to the operating system kernel or vendor-provided code. The kernel and the driver on the \jetsoncomplet are open-source. However such modifications require costly maintenance to be kept up-to-date with the upstream code. 
\end{requirement}

\begin{requirement}[Maintenance]
\label{requirement_resiliency}
The method should not be tied to a specific revision of the software platform, or offer a clear path for revisions. Updates to the software platform tend to be backwards compatible, allowing older software or methods to work. Methods which rely on deprecated or undocumented behaviours risk becoming outdated. 
\end{requirement}

\begin{requirement}[Scope]
\label{requirement_scope}
The access control shall focus on interference between operations running on the \gpu issued by different applications and their latency. Interference between the CPU and the \gpu, or stemming from other accelerators on the platform are outside the scope of this work.
\end{requirement}

\begin{requirement}[Temporal]
\label{requirement_temporal}
The access control shall focus on temporal access control techniques, for the \gpu as a whole. 
The \jetson \gpu offers no hardware support or guarantees on the isolation of its resources. 
\end{requirement}

\begin{requirement}[Burst preservation]
\label{requirement_burst}
The access control shall maintain existing synchronisation barriers. The resource manager may introduce new synchronisation barriers, splitting an existing burst into one or more bursts.
This is required to maintain the correctness of executed bursts. Additional synchronisation points may provide some flexibility without jeopardising the functional correctness of applications.
\end{requirement}

\begin{requirement}[Order preservation]
\label{requirement_ordering}
The access control shall maintain the relative ordering between \gpu operations. The access control may introduce stronger guarantees on the ordering between \gpu operations. As per Aspect \ref{requirement_burst}.
\end{requirement}

\begin{requirement}[Throttling]
The access control defers the insertion of \gpu operations into the CUDA software stack to improve isolation between applications. Once operations enter the CUDA Software stack, and in particular streams, their execution is deferred to the platform, and only limited control and guarantees are available.
\end{requirement}

\begin{requirement}[Software]
The access control shall rely on user-level, software-only methods. As per Aspect \ref{requirement_temporal}, hardware and software support on the platform is limited.
\end{requirement}

%% file: sec_resource_manager.tex

\section{COOK Access Controller}
\label{sec_resource_manager}

We introduce in the following an access control technique based on deferring \textit{when} \gpu operations enter \gpu streams to control their execution, effectively throttling the flow of \gpu operations to the \gpu and its driver.
Our approach relies on the configurable generation of C hooks (COOK) to alter the behaviour of \gpu routines (Section \ref{sec_rm_hooks}). 
We focus on hooks for the CUDA Runtime library, which all applications will call, directly or indirectly, to schedule operations on the \gpu.
While the driver and driver API would be appropriate candidates for hooking, their interfaces are not as well documented and are more volatile.
We propose policies to enforce access control in line with our requirements in Section \ref{sec_rm_strategies}.

The definition of a configurable process is justified by Aspect~\ref{requirement_resiliency}. New versions of the software platform may deprecate, add to, remove from, or alter the behaviour of the hooked library. Neither deprecation nor removal should alter the COOK configuration or the generated hooks.
At best, the same configuration may apply to multiple versions of the hooked library. 
Additions and alterations need to be considered separately to assess whether the existing hook templates remain applicable.
The configuration may be revised, with new conditions, but still rely on existing hook templates.
At worst, a whole new configuration is required. 

\subsection{Hook generation}
\label{sec_rm_hooks}

\begin{figure}[htb]
	\centering
	\includegraphics[width=0.65\linewidth]{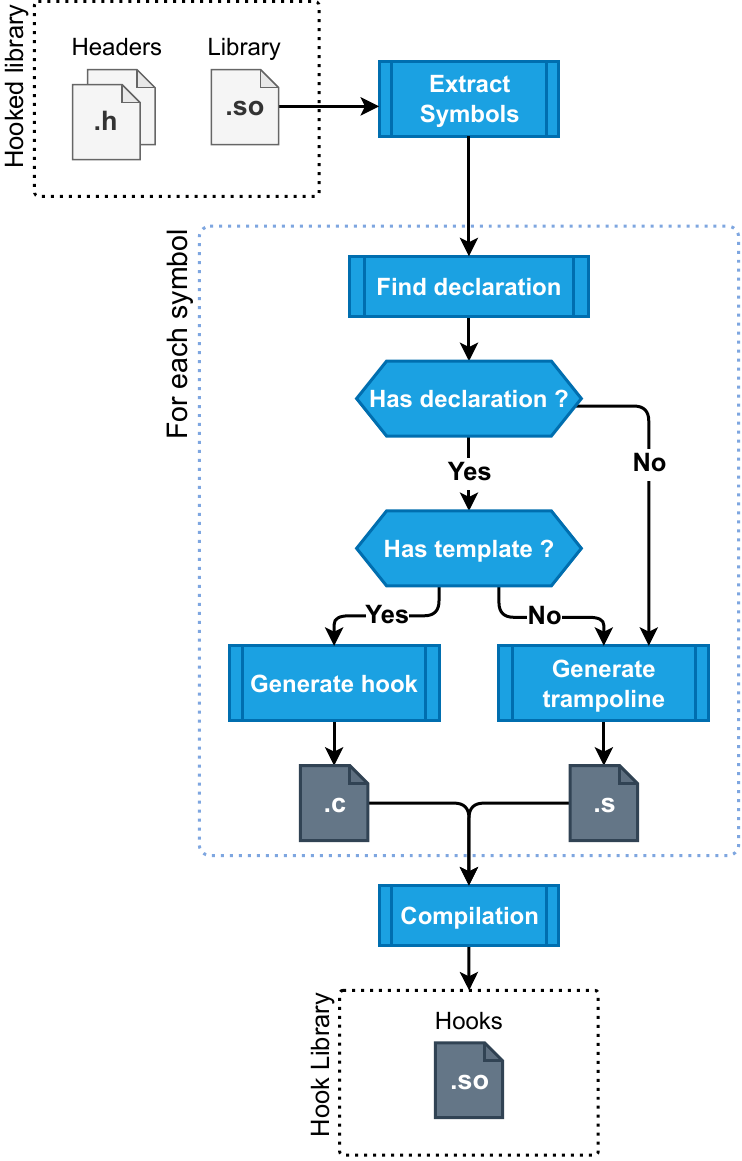}
	\caption{Generation of a hook library
                 \label{fig-cook_generation}}
\end{figure}

The COOK toolchain generates a hook library to intercept all calls to a selected hooked library. Specifically, we hook onto the CUDA Runtime library, \textit{libcudart.so}\footnote{This contribution focuses on dynamic libraries without loss of generality. Static library hooks are generated in a similar fashion.}, as it is on the critical path to the \gpu for the CUDA software stack.
The generation of the hook library is a process configured by a set of hook templates and conditions.
Each hook template is a code template instantiated with a function declaration to create a corresponding hook. Hook conditions capture the list of functions to hook onto for each template.
The overall workflow of the COOK toolchain for a given library is outlined in Figure~\ref{fig-cook_generation}. It is as follows:
\begin{itemize}
\item \emph{Extract symbols}: list the symbols exported by the hooked library to capture its interface;
\item \emph{Find symbol declaration}: for each symbol $s$, find its signature in the library headers to capture its arguments and their types;
\item \emph{Generate a hook}: for each symbol $s$, if a hook condition matches then apply its template to generate a hook, i.e. a function $s^h$ that intercept calls to $s$;
\item \emph{Generate a trampoline}: for each symbol $s$ undefined or condition-less, generate a default, minimal hook $s^h$ configured to simply raises an error, or simply redirect calls to the hooked function $s$~\cite{implib}; 
\item \emph{Compilation}: gather all generated hooks and trampolines and compile the hook library.
\end{itemize}

\begin{figure}[htb]
	\centering
	\includegraphics[width=0.65\linewidth]{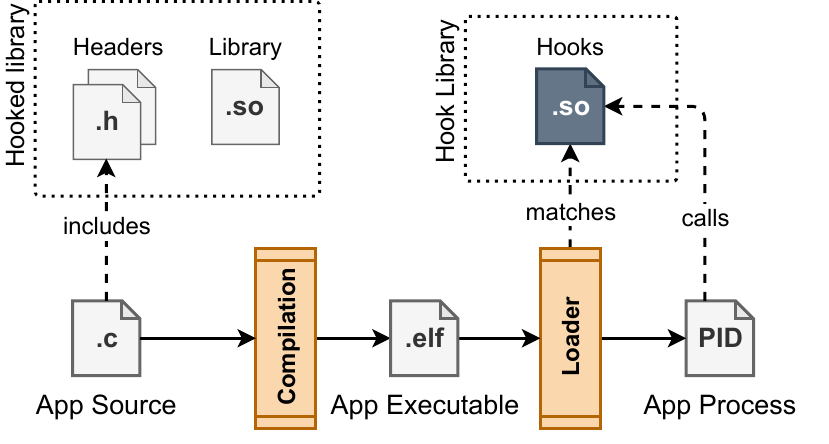}
	\caption{Use of a hook library by an application
                 \label{fig-cook_usage}}
\end{figure}

During compilation (Figure~\ref{fig-cook_usage}), an application may rely on symbol declarations provided by library headers to generate calls, listing undefined symbols. 
Undefined symbols are then resolved at runtime by the loader which match the libraries required by the application with available ones.
A match must have the correct name, e.g. \textit{libcudart.so}, and expose all symbols required for the execution of the application executable.
The hook library could expose only the hooked symbols, without trampolines, with the loader resolving potential gaps. 
In practice, however, some CUDA libraries and applications may circumvent the loader to load the CUDA Runtime, looking instead into specific paths. 
As such, the generated hook library must be able to replace the CUDA Runtime in-place with all its symbols (Aspect~\ref{requirement_transparency}).

\subsection{Access control strategies}
\label{sec_rm_strategies}

We propose three access control strategies to mitigate the impact of interference on \gpu operations scheduled by concurrent applications. They effectively aim to schedule GPU operations in a non-preemptive
fashion. 
We depict for each strategy how it hooks related \gpu routines. All strategies rely on the same principles:
\begin{itemize}
\item Any operation running on the \gpu should have exclusive use of the \gpu resources (Aspect \ref{requirement_temporal}). This is achieved by ensuring an operation only executes if its application has first acquired the global \gpu lock \texttt{GPU\_LOCK}. Only one application can \texttt{acquire} the lock at any given time. Calls to \texttt{acquire} from other applications will be blocked until a \texttt{release} from the current owner of the \gpu lock\footnote{Our implementation uses a semaphore from the POSIX threads library, and the underlying scheduling policy, included as part of the \jetsoncomplet platform. The election of an operation amongst all pending candidates for execution, that is the definition of a \gpu-specific scheduler, is an orthogonal problem outside the scope of this work.}.
\item Strategies hook into the \gpu routines generating \emph{copy} and \emph{kernel} \gpu operations, respectively the \texttt{cudaLaunchKernel} and  the \texttt{cudaMemcpy} methods in the CUDA runtime API. Each routine inserts the corresponding operation and its parameters into the application stream as depicted in Algorithms~\ref{alg-cuda_launch} and~\ref{alg-cuda_memcpy}.
\end{itemize}
%






\begin{algorithm}
\caption{Kernel Launch method in the CUDA Runtime}\label{alg-cuda_launch}
\begin{algorithmic}[1]
\small
\Procedure{cudaLaunchKernel}{$func, grid, args, stream$}
\State \textbf{insert op} Execute(func, grid, args) \textbf{in} stream
\EndProcedure
\end{algorithmic}
\end{algorithm}

\begin{algorithm}
\caption{Memory copy method in the CUDA Runtime}\label{alg-cuda_memcpy}
\begin{algorithmic}[1]
\small
\Procedure{cudaMemcpy}{$dst, src, size, mode, stream$} 
\State \textbf{insert op} Copy(dst, src, size, mode) \textbf{in} stream
\EndProcedure
\end{algorithmic}
\end{algorithm}


\subsubsection{Host Callback (callback) strategy}
\label{sec_strategy_callback}
\begin{figure}[htb]
	\centering
	\includegraphics[width=0.55\linewidth]{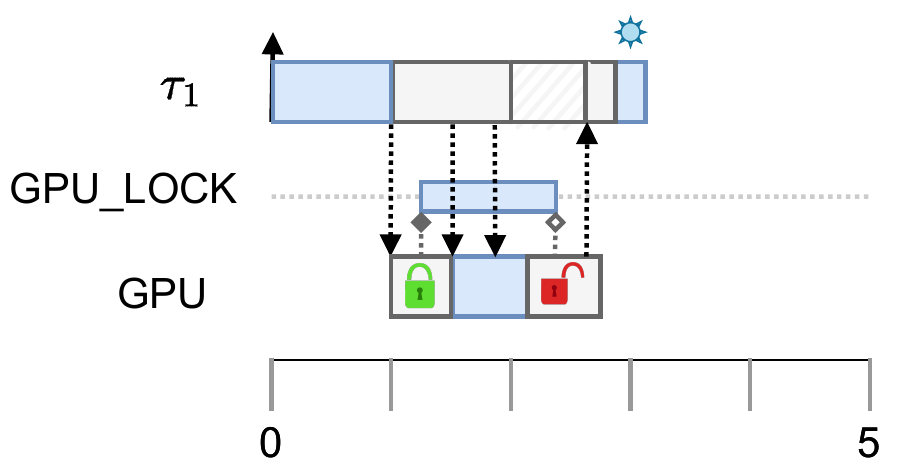}
	\caption{Illustration of the callback strategy
                 \label{fig-cook_callback}}
\end{figure}

The callback strategy relies on the execution order guaranteed by a \gpu stream, i.e. operations inserted in a \gpu stream are executed in a First-In First-Out order.
An operation cannot start until all the previous ones completed.
As illustrated in Figure \ref{fig-cook_callback}  (hook-related changes use a lighter shade), the callback strategy adds an \texttt{acquire} operation (first block on the \gpu) before any \gpu operation (second block on the \gpu) with a matching \texttt{release} afterwards (last block on the \gpu).

The operation acquiring the \gpu mutex will thus block the execution of operations in its stream until the \gpu is available.
Conversely, the release operation will only free the \gpu once all prior operations complete.
Only one stream across all applications can proceed through the \gpu lock at any time. Others are either ineligible for scheduling by the \gpu, or waiting on a blocking \texttt{acquire} operation.
The approach uses the CUDA Runtime \texttt{cudaLaunchHostFunc} to revert control back to the CPU to execute a specific host method (\texttt{acquire} and \texttt{release}). 
Algorithm~\ref{alg-cuda_launch-callback} outlines the generated hook for the kernel launch method. 
(The memory copy uses the same code template resulting in a similar hook.)

\begin{algorithm}
\caption{Kernel Launch hook (callback strategy)}\label{alg-cuda_launch-callback}
\begin{algorithmic}[1]
\small
\Procedure{CookLaunchKernel}{$func, grid, args, stream$}
\State \textbf{insert op} Callback(\texttt{acquire} GPU\_LOCK) \textbf{in} stream
\State \textbf{insert op} Execute(func, grid, args) \textbf{in} stream
\State \textbf{insert op} Callback(\texttt{release} GPU\_LOCK) \textbf{in} stream
\EndProcedure
\end{algorithmic}
\end{algorithm}

\subsubsection{Synchronised Operation (synced) strategy}
\label{sec_strategy_synchronised}

\begin{figure}[htb]
	\centering
	\includegraphics[width=0.55\linewidth]{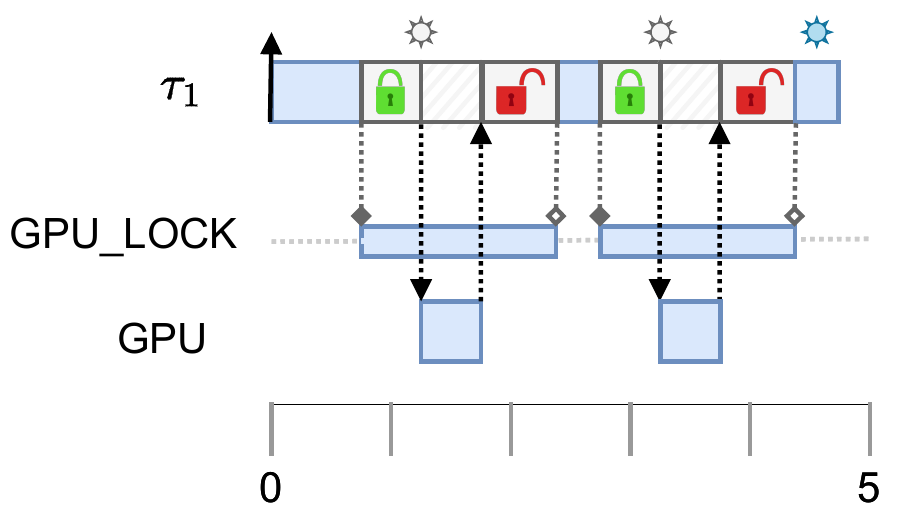}

	\caption{Illustration of the synced strategy
                 \label{fig-cook_synced}}
\end{figure}

The synced strategy transforms \gpu routines into synchronisation points. 
A call to a routine only completes after the related \gpu operation does. This is comparable to the RGEM \cite{rm:rgem} approach.
Figure \ref{fig-cook_synced} shows the behaviour of the hook. 
The hook first acquires the \gpu lock (downward arrow to the lock), before actually calling the \gpu routine (downward arrow to the \gpu). The hook then waits for the operation's completion (star symbol). Once the operation is complete (upward arrow from the \gpu), the hook releases the lock (downward arrow to the lock).
An application thus schedules and executes at most one \gpu operation at a time (thanks to the barrier).
 And only one application can schedule a \gpu operation at any time (thanks to the lock).
Algorithm~\ref{alg-cuda_launch-synchronised} outlines the generated hook for the kernel launch method. 
(Like the prior strategy, the memory copy uses the same code template resulting in a similar hook.)

\begin{algorithm}
\caption{Kernel Launch hook under the synced strategy}\label{alg-cuda_launch-synchronised}
\begin{algorithmic}[1]
\small
\Procedure{CookLaunchKernel}{$func, grid, args, stream$}
\State \texttt{acquire} GPU\_LOCK
\State \textbf{insert op} Execute(func, grid, args) \textbf{in} stream
\State \textbf{sync on} device 
\State \texttt{release} GPU\_LOCK
\EndProcedure
\end{algorithmic}
\end{algorithm}

\subsubsection{Deferred Worker (worker) strategy}
\label{sec_strategy_worker}

\begin{figure}[htb]
	\centering
	\includegraphics[width=0.55\linewidth]{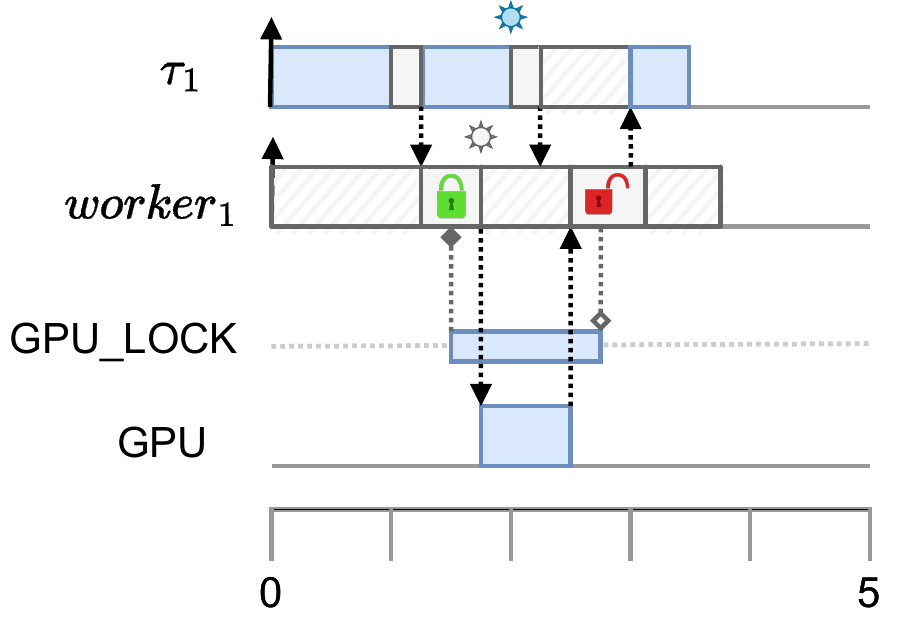}
	\caption{Illustration of the worker strategy
                 \label{fig-cook_worker}}
\end{figure}

The worker strategy defers the execution of \gpu routines and operations to a separate \emph{worker} thread running on a separate CPU core for each application. As illustrated in Figure \ref{fig-cook_worker}, the hooked \gpu routine call from the task (downward arrow) notifies the worker instead of the \gpu.
Then for each application, and similarly to the synced strategy, its \emph{worker} acquires the \gpu lock (downward arrow to the lock), 
queues the \gpu operation 
for execution (first downward arrow to the \gpu), 
waits for the operation to complete (upward arrow from the \gpu), and releases the lock (last downward arrow to the lock).
The \gpu operation effectively transits through the worker (using a new $worker\_queue$ stream per worker) to the \gpu.
Each worker acts as a synchronisation source, ensuring a synchronisation barrier only releases an application once the worker completed all queued work (upward arrow from the worker).
Workers from different applications synchronise to ensure only one of them schedules an operation at any given time.

Algorithm~\ref{alg-cuda_launch-worker} outlines the generated hook for the kernel launch method. (Like the prior policies, the memory copy uses the same code template resulting in a similar hook.) The kernel launch operations and its parameters are copied into the $worker\_queue$ instead of the designated stream.
The worker behaviour is described in Algorithm~\ref{alg-worker-worker}. It loops over the $worker\_queue$ and operates by dequeuing an operation, then executing it, and waiting for its completion. The execution first requires the worker to acquire the \gpu lock, released once the operation completes. \todo{Add algorithm lines number in description}

\begin{algorithm}
\caption{Kernel Launch hook under the worker strategy}\label{alg-cuda_launch-worker}
\begin{algorithmic}[1]
\small
\Procedure{CookLaunchKernel}{$func, grid, args, stream$}
\State \textbf{insert op} Execute(func, grid, args) \textbf{in} worker\_queue
\EndProcedure
\end{algorithmic}
\end{algorithm}

\begin{algorithm}
\caption{Worker thread implementation}\label{alg-worker-worker}
\begin{algorithmic}[1]
\small
\Procedure{CookWorkerThread}{}
\While{true}
	\State $op \gets head(worker\_queue)$
	\State $pop(worker\_queue)$
	\If{$op$ \textbf{is} Execute(...)}
		\State \texttt{acquire} GPU\_LOCK
		\State \textbf{insert op} Execute(...) \textbf{in} stream
		\State \textbf{sync on} stream 
		\State \texttt{release} GPU\_LOCK
	\ElsIf{$op$ \textbf{is} Copy(...)}
		\State \texttt{acquire} GPU\_LOCK
		\State \textbf{insert op} Copy(...) \textbf{in} stream
		\State \textbf{sync on} stream 
		\State \texttt{release} GPU\_LOCK
	\EndIf
\EndWhile
\EndProcedure
\end{algorithmic}
\end{algorithm}

A burst of operations may contain operations other than kernel launches and memory copies, e.g. host callbacks. Those still need to be executed in a FIFO order to maintain the semantic and behaviour of the application (Aspect \ref{requirement_ordering}). 
The worker could support all stream-ordered operations, however this would increase its complexity. 
Instead, we focus on the \emph{kernel} and \emph{copy} \gpu operation types.
Other operations must synchronise with the worker, ensuring it has completed all prior operations in the burst before proceeding themselves. This is achieved through the shared code template in Algorithm~\ref{alg-worker-others}.

\begin{algorithm}
\caption{Operations hook under the worker strategy}\label{alg-worker-others}
\begin{algorithmic}[1]
\small
\Procedure{CookOrderedOp}{$args...$}
\State \textbf{sync on} $worker\_stream$ 
\State \textbf{insert op} Op(args...) \textbf{in} stream 
\EndProcedure
\end{algorithmic}
\end{algorithm}

Note that upon a kernel launch, we need to copy the list of arguments passed to the kernel for its deferred execution by the worker. 
This argument list may be allocated on the stack, as is the case for code generated by the \gpu compiler. The list may thus have been deallocated when the actual execution occurs.
To create the argument copy, our implementation references a list of known kernels in the application.
For each kernel, the list holds the number of parameters it requires, their size, and the memory layout of the argument list. 
The worker strategy currently intercepts calls to the CUDA Runtime kernel registration primitives, \texttt{\_\_cudaRegisterFunction}, to create said list.
The kernel registration primitives are used by the CUDA runtime to register information about kernels on the host side, notably the kernel name, memory address, and binary code.
The kernel registration primitives are undocumented functions of the CUDA Runtime, thus our implementation may not be resilient to CUDA Runtime updates, failing to adhere to Aspect~\ref{requirement_resiliency}. 
However, the principles of the strategy still hold. The kernel list could be built through an off-line analysis of the application.

%% file: sec_methodology.tex

\section{Methodology}
\label{sec_methodology}


We aim to evaluate the impact of interference resulting from different applications sharing the \gpu, the benefits of temporal access control to mitigate said interference, and the complexity of the hooks.
This section presents the methodology used in our evaluation: the considered platform, instrumentation for kernel measurements, collected metrics, and application configurations.


\subsection{Platform Configuration}
\label{sec_metho_platform}

All experiments were collected on a \jetsoncomplet setup with Jetpack version 5.0.2. CUDA version 11.4 was used for all benchmarks.  The platform is running Ubuntu 20.04.6 LTS (Kernel 5.10.104-tegra). It is configured under the \emph{MAXN} power profile, all cores are active, and the CPU and \gpu frequency are allowed to vary, respectively in the 114MHz-1.3GHz and 1.19GHz-2.27GHz ranges, in response to workload changes or throttling constraints.

\subsection{Instrumentation}
\label{sec_metho_instrumentation}

We instrument the executed kernels to monitor their behaviour. We consider two methods for instrumentation, application- and kernel-level. 
Application-level instrumentation collects a trace of CUDA calls, using the \nvidia \emph{nsys} tracing solution, a high-level view of the application execution. \emph{nsys} probes into unmodified applications.
Kernel-level instrumentation traces the end-to-end execution of each thread block, using our own instrumentation primitives, for an overview of how each kernel is dispatched and executed. Applications, where possible, are modified to call the instrumentation routines and collect traces.
Instrumentation, in spite of our best efforts, will introduce noise in the measurements. However, all runs should suffer comparable noise, under a given instrumentation technique. 


\subsection{Benchmarks}
\label{sec_metho_bench}

We consider two types of benchmarks: 


\begin{itemize}
\item \benchmmult is an ad-hoc CUDA benchmark, with explicit kernel definitions and calls to the CUDA Runtime. It is provided as a sample of CUDA code by \nvidia. The benchmark is composed of a single burst which repeatedly calls the same matrix multiplication kernel~($300\times$) over the same input data. Measurements are collected for a single run of the benchmark.
\item \benchdna is a model-based benchmark, using the ONNX runtime \cite{onnx} to schedule  a DNN model and offload computation to the \gpu. It is an industrial case study which performs drone detection and avoidance. Each inference is composed of long bursts with few synchronisation points. Input data is randomly generated for each inference. Measurements are collected over a 60s sampling period after a 30s warm-up time.
\end{itemize}

\subsection{Configurations}
\label{sec_metho_configurations}

We consider each application under different configurations to assess the impact of interference, and that of each hook strategy. A configuration is denoted as such:
\begin{center}
\benchcfg{bench}{isol}{strategy}
\end{center}

Where \texttt{bench} is one of the benchmarks identified in Section~\ref{sec_metho_bench}. \texttt{isol} is one of the following, identifying if the benchmark is running in parallel:
\begin{itemize}
\item isolation: the benchmark is running in isolation;
\item parallel: 2 instances of the benchmark application are running in parallel (mirrored).
\end{itemize}

The \texttt{strategy} modifier indicates the hook strategy used when running the application (or applications): \texttt{none} (No hook), \texttt{callback}, \texttt{synced}, and \texttt{worker}.
The \benchcfg{\benchmmult}{parallel}{synced} configuration is thus the \benchmmult benchmark running in parallel with itself under the synchronised operation strategy.

\subsection{Metrics}
\label{sec_metho_metrics}

Our evaluation aims to assess a number of aspects, notably the impact of the proposed strategies on interference between \gpu operations, their impact on the performance of an application in isolation, and their complexity in terms of deployment and maintenance. To that end, we collect and present the following metrics:
\begin{itemize}
\item The Normalised Kernel runtime (NET) captures the variation in the execution time of kernels. Large NET ranges indicate respectively high execution time variability, inherent to the kernel or due to interference on shared resources.
$NET_{k,c}^i$ for the i$^{th}$ instance of a kerne~($k$) under a given configuration ($c$) is computed as the ratio between the observed execution time of the kernel ($ET_{k,c}^i$) and the lowest observed execution time of the kernel under the same configuration:
\begin{equation} \label{eq-metric_net}
NET_{k,c}^i = \frac{ET_{k,c}^i}{min_j(ET_{k,c}^j)}
\end{equation}

\item The Inferences per Second (IPS) is a performance metric for applications. Low IPS indicate a slow-running application, and comparing IPS in isolation and in parallel is indicative of the slow down suffered by the application due to interference.
$IPS_{a,c}^t$ is computed as the ratio between the number of completed executions ($N$) of the application ($a$) under a given configuration ($c$) in the measurement interval ($t$). It is measured by looping over the application under randomised or fixed inputs, counting completed executions at regular intervals (1s). 
\begin{equation} \label{eq-metric_net}
IPS_{a,c}^t = \frac{N_{a,c}^t}{duration(t)}
\end{equation}
 
\item The number of Lines of Code (LoC) Lines of Code is a metric of code complexity. High LoC is indicative of a large code base and correlates to high development and maintenance costs for an application or a tool. LoC is measured for configuration, hook template, and generated code files using \emph{cloc} \cite{cloc}.
\end{itemize}


%% file: sec_evaluation.tex

\section{Evaluation}
\label{sec_evaluation}


\subsection{Assessing the impact of interference}
\label{sec_evaluation_interference}

\begin{figure*}[hbt]
	\centering
	\includegraphics[width=\textwidth]{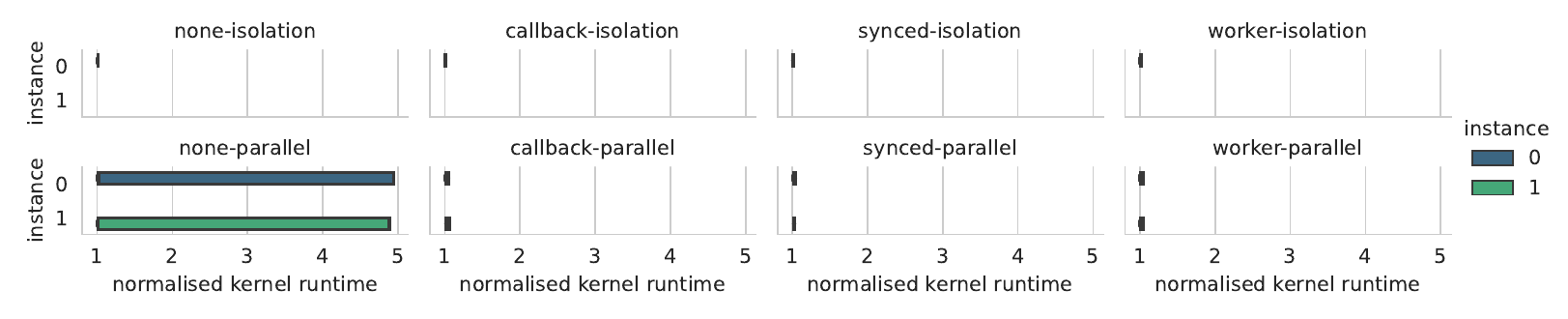}
	\caption{Distribution of normalised kernel runtimes for the \benchmmult benchmark under all configurations
		\label{fig-net_mmult}
		\label{fig-net_mmult_cook}}
\end{figure*}

\begin{figure*}[hbt]
	\centering
	\includegraphics[width=\textwidth]{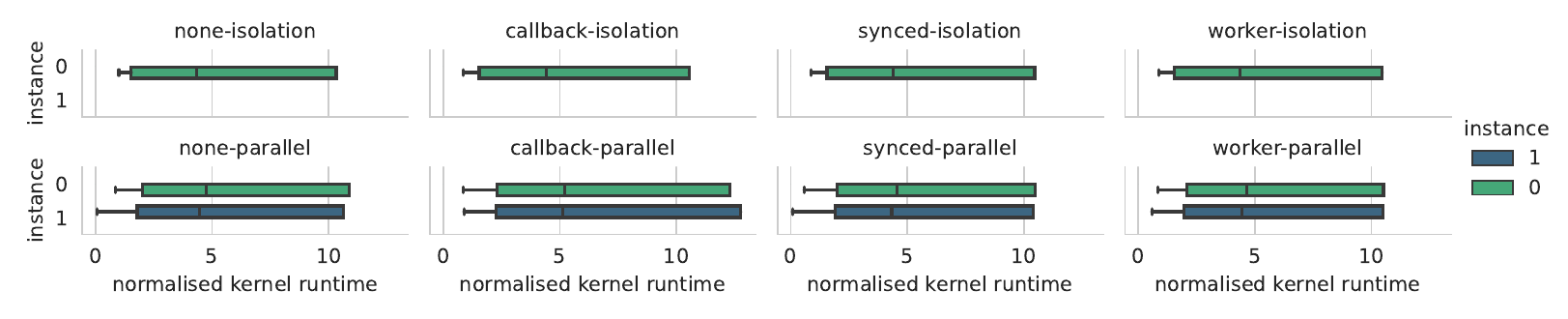}
	\caption{Distribution of normalised kernel runtimes for the \benchdna benchmark under all configurations
		\label{fig-net_dna}
		\label{fig-net_dna_cook}}
\end{figure*}

We first assess the impact of interference on kernels from different applications running on the \gpu. To that end, we compare the Normalised Kernel Runtime (NET) for each benchmark \texttt{bench} in isolation (\benchcfg{bench}{isolation}{none}) and in parallel (\benchcfg{bench}{parallel}{none}), measured with the \textit{nsys} instrumentation tool. 
The results are presented in Figures~\ref{fig-net_mmult} to \ref{fig-net_dna} respectively for \benchmmult and \benchdna. 
Each Figure captures the distribution of NET for each instance of the benchmark across all executed kernels  as a boxplot. 
The box captures the $50$\% of the samples around the median, the whiskers capture 99\% of the data, and outliers in the lowest and highest $0.5\%$ have been omitted for readability.

All benchmarks follow a similar trend where their performance in parallel is impeded by  sharing of the \gpu with another application. 
Interference does result in higher kernel execution time variability, as identified by a larger range of values.
Some benchmarks, e.g. \benchdna, exhibit an inherent variability of kernel execution times but minor effects of interference on the majority of kernel calls.
Interference also results in occasionally large slowdowns, \slowdown{5.5} for \benchmmult and, very rarely  \slowdown{1200} for \benchdna (less than $0.5\%$ of kernels exceed a $10\times$ slowdown).
Outliers for \benchmmult never exceed a \slowdown{5.5} slowdown, and all policies reduce their occurrence such that less than $0.5\%$ of kernels suffer so. 
The policies also affect outliers for \benchdna, reducing the maximum observed slowdown from \slowdown{1200} (\benchcfg{\benchdna}{parallel}{none} and\benchcfg{\benchdna}{parallel}{callback}) to the same as in isolation around \slowdown{200} (respectively \slowdown{800}) under \benchcfg{\benchdna}{parallel}{synced} (respectively \benchcfg{\benchdna}{parallel}{worker}).


\begin{table}[htb]
\center
\begin{tabular}{ |l|rrrr| } 
 \hline
Configuration  & none & callback & synced & worker \\ \hline
isolation  & 113 & 37 & 67 & 84 \\
parallel   & 49 &  32  & 25 & 26 \\
 \hline
\end{tabular}
\caption{Inference per Second (IPS) achieved by the onnx\_dna benchmark for the different configurations
			\label{tab-cook_ips_dna}}
\end{table}

In terms of overall application performance, we consider the IPS achieved by the application, and in particular of \benchdna in Table~\ref{tab-cook_ips_dna}  (results include all configurations, including the mitigation strategies). In isolation (\benchcfg{\benchdna}{isolation}{none}), the application achieves an IPS of 113. This falls down to 49 in the matching parallel configuration (\benchcfg{\benchdna}{parallel}{none}).
While a decrease of the IPS could be expected as two instances share the \gpu, it is still slightly more than \slowdown{2} for only 2 parallel instances. It however does not reflect the highest slowdown suffered by individual kernel calls.

\subsection{Assessing the impact of mitigation}
\label{sec_evaluation_mitigation}

\begin{figure*}[htb]
	\centering
\begin{subfigure}{0.19\textwidth}
	\includegraphics[width=0.7\textwidth]{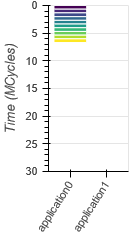}
    \caption{\benchcfg{}{isolation}{none}}
\end{subfigure}
\hfill
\begin{subfigure}{0.19\textwidth}
	\includegraphics[width=0.7\textwidth]{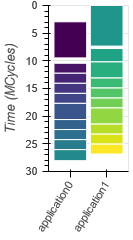}
    \caption{\benchcfg{}{parallel}{none}}
\end{subfigure}
\hfill
\begin{subfigure}{0.19\textwidth}
	\includegraphics[width=0.7\textwidth]{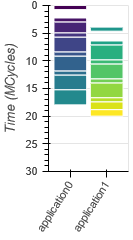}
    \caption{\benchcfg{}{parallel}{callback}}
\end{subfigure}
\hfill
\begin{subfigure}{0.19\textwidth}
	\includegraphics[width=0.7\textwidth]{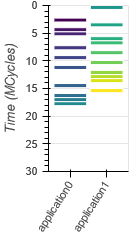}
    \caption{\benchcfg{}{parallel}{synced}}
\end{subfigure}
\hfill
\begin{subfigure}{0.19\textwidth}
	\includegraphics[width=0.7\textwidth]{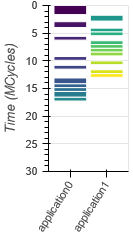}
    \caption{\benchcfg{}{parallel}{worker}}
\end{subfigure}
	\caption{Chronogram of the \benchmmult benchmark execution under the various configurations
                 \label{fig-chronogram_mmult}
                 \label{fig-chronogram_mmult_cook}
                 }
\end{figure*}

We now assess the impact of the proposed mitigation strategies on the execution of parallel applications. We focus in particular on ensuring the proposed strategies do manage to isolate the execution of \gpu operations from different applications. To that end, we instrument the kernel of the \benchmmult benchmark to trace the execution of each executed kernel on the \gpu. The results are presented as chronograms in Figure~\ref{fig-chronogram_mmult}. Each chronogram captures the trace of kernel executions, from the beginning of their first executed block (top) to the completion of their last (bottom). 
Different columns indicate blocks belonging to different instances of the same benchmark.

Figure~\ref{fig-chronogram_mmult} first compares the execution of \benchmmult in isolation and in parallel without any hooked library. 
We first note that the application suffers a \slowdown{4} slowdown due to sharing the \gpu, taking about 28 Million Cycles to complete over 8 in isolation. 
This is similar to the slowdown observed in the previous section on normalised kernel runtimes.
Some kernels appear to take much longer when their execution overlaps. 
In practice, the \jetsoncomplet does not allow two applications (and two \gpu contexts) to run concurrently; it constantly switches contexts to allow both instances to progress without starvation.
As was illustrated in Figure~\ref{fig:diag_burst}, the unknown behaviour of the platform in parallel does not allow us to highlight when and how often such switches occur.
Context switches are costly on the \gpu as all registers of all SM may need to be saved to memory. Kernels from an application further suffer from cache-related preemption delays as they have to reload useful blocks in the caches evicted by the other.


We also consider the impact of the hook strategies in terms of mitigation in Figure~\ref{fig-chronogram_mmult_cook}. 
Both the synced and worker strategies do manage to isolate the execution of kernels from different instances of \benchmmult, with no visible overlap between their blocks.
The callback strategy however fails to isolate \gpu operations (host callbacks possibly cause a \gpu context switch but none of the cache polluting effects). 
All strategies outperform the configuration without mitigation, and they achieve a similar performance with a slight benefit for the worker one.
All strategies also outperform a PTB solution (using~\cite{rm:FGPU}, omitted for the sake of brevity), where both instances were allocated 4 \gpu SMs. The PTB implementation did rely on modifications of the application (Aspect~\ref{requirement_transparency}).
The benchmark still suffers a slowdown greater than the number of running instances.
The impact of the hook strategies on performance is considered in the next section.

\subsection{Impact of the hook strategies}
\label{sec_evaluation_hook}

We now consider the impact of the various hook strategies to mitigate the interference observed in Section~\ref{sec_evaluation_interference}. We repeat the same experiment running each \texttt{bench} in isolation and in parallel configurations, but with all instances running under one of the hook \texttt{strategy}. The results are presented in Figures~\ref{fig-net_mmult_cook} to \ref{fig-net_dna_cook} respectively for \benchmmult and \benchdna. The top row captures configurations in isolation, while the bottom one is for benchmark in parallel ones. Each column holds the result for a specific strategy, from left to right: none, callback, synced, and worker.

All benchmarks again follow a similar trend. They still suffer from long running kernels, with worst-case slowdowns similar to the ones observed without mitigation: \slowdown{5.5} for \benchmmult and \slowdown{1200} for \benchdna. However these slowdowns are less frequent under all hook strategies, as is most notable for the \benchmmult benchmarks where 99\% of executed kernels suffer negligible slowdowns. 
There is little difference overall in the impact of the various strategies for \benchmmult.
For \benchdna, the callback strategy does increase the variability of kernel execution times.
Conversely, both the synced and worker strategies manage to reduce the variability introduced by the second instance of the benchmark 
and the maximum observed slowdown respectively to $200\times$ (close to the maximum in isolation) and $800\times$.
The synced and worker strategies do isolate the execution of individual \gpu operations (see Section~\ref{sec_evaluation_mitigation}).
The high slowdowns hint that sharing the \gpu during the lifetime of applications even at a high granularity still results in slowdowns.
Interference occurs outside the execution of \gpu operations, e.g. in shared software resources, or due to cache pollution or costly context switches.

We assess the overall performance of the \benchdna application using the IPS metric in Table~\ref{tab-cook_ips_dna}. Like Figure \ref{fig-net_dna_cook}, the top row holds the information for the isolation configurations, and each column is a strategy. 
The various strategies do result in significant slowdown for the benchmark, in isolation and in parallel. The hook strategies result in worse performance than the unmitigated scenario.
This may be due to the additional synchronisation between applications and a reduced use of the \gpu resources within each instance of the application.

All strategies do introduce additional synchronisation points in the application, potentially after each \gpu operation (callback and synced).
The callbacks further add a considerable overhead to the application execution.
This breaks up the long bursts of \benchdna into smaller ones thus slowing the preparation and scheduling of \gpu operations on the host-side. 
The worker strategy is the less invasive one, as it allows the application to proceed while it schedules synchronise \gpu operations itself. However, unmanaged \gpu operations (outside copy and execution) still need to synchronise with the worker to guarantee the overall order of operations scheduled on the \gpu.

\subsection{Complexity of the hook generation}
\label{sec_evaluation_complexity}

We assess in the following the complexity of the hook strategies, as a proxy for the resiliency of the method. Simple methods or strategies shall be easier to maintain over time. 
We first present in Table~\ref{tab-cook_loc} the Lines of Code for different artefacts of the toolchain, namely the configuration file and templates used to generate the hooked library. In the absence of hook generation, that code would have to be maintained in whole.

\begin{table}[htb]
\center
\begin{tabular}{ |l|rr|r| } 
 \hline
Strategy  & Configuration & Templates & Generated code \\ \hline

callback & 153 & 151 & 6804 \\
synced & 153 & 149 & 6813 \\
worker & 171 & 1056 & 8383 \\
 \hline
\end{tabular}
\caption{Lines of Code (LoC) required and generated for the different strategies
			\label{tab-cook_loc}}
\end{table}

All strategies rely on small configuration files, mostly defining symbols and the templates to generate the related hook definition. The templates LoC accounts for both the common code for the strategy, as well as symbol-specific ones. 
The worker strategy is the more complex one in both aspects.
Additional templates apply to \gpu operations managed by or synchronised with the worker queue, thus requiring additional entries in the configuration. 
The measure also includes all the code related to the deferred worker execution and worker queue management. 
Overall, this represents at most a thousand hundred lines of code to maintain. But the hook generation results in thousands of lines of code.

We configured the tool to raise an error on calls to all CUDA Runtime methods by default, unless explicitly hooked or ignored.
This ensures that an application cannot call methods which may generate unmanaged \gpu operations. Such methods need further consideration and an appropriate hook definition.
We further distinguish between \emph{implicit} and \emph{unknown} symbols in this category.
\emph{Implicit} symbols have no explicit hook or exclusion rule.
 \emph{Unknown} symbols are ones for which the tool could find no matching declaration.
In practice, unknown symbols are not used in our benchmarks. They correspond to variants of CUDA methods where the default CUDA stream is not shared within a \gpu context. Their declaration is generated in the CUDA Runtime headers based on some compilation-time configuration. 
We have considered workarounds to retrieve the generated definition, e.g. pre-process said headers, to circumvent it, e.g. the trampoline could call C pre- and post hooks, or to map the variants to the original declaration. But we have yet to explore them.

Considering symbols hooked by trampolines, 
the hook strategies intercept less than 70 methods of the CUDA Runtime library, out of 385.
Those are mostly variants of copy (\texttt{cudaMemcpy}) and execution (\texttt{cudaKernelLaunch}) \gpu routines.
The worker strategy in addition distinguishes methods which create or depend on synchronisation points to ensure these also synchronise with the worker queue. The same template is used for all of them.
While this requires a separate assessment, as to which methods should be included or not in the hooked methods, it only needs to be performed once and then simply updated upon new versions of the library.

%

%% file: sec_conclusion.tex

\section{Conclusion}
\label{sec_conclusion}

The \nvidia~\jetsoncomplet can deliver high performance and throughput
in a small SWaP package, by combining general purpose and specialised cores.
However, interference, especially on its \gpu, does result in significant slowdowns and execution time variability.
The problem is compounded by the closed nature of the platform.
We considered the specification and implementation of an access controller to mitigate the interference suffered by applications sharing the \jetson \gpu.
From the state of the art, we outlined a number of aspects to minimise the impact on applications w.r.t. deployment and development. 
A second concern was to ease the deployment and maintenance of the method across variants of the platform.



Our approach focuses on ensuring \gpu operations from separate applications have exclusive access to the \gpu during execution.
We proposed various strategies reliant on intercepting \gpu operations calls through hooks on the \gpu routines.
All strategies manage to reduce the impact of \gpu interference on operations themselves, with two strategies achieving  isolation.
Benchmarks such as the ONNX runtime, which benefit from the CPU and the \gpu working in tandem, do suffer a performance loss at the application from the additional synchronisation barriers introduced by our approach. 
The use of a generative approach to generate hooks allows for a small, manageable implementation for all strategies. 
Changes to the software stack can be accounted for by adding new hook templates or reusing existing ones.

There is still work to be done in the context of interference within the \gpu.
Even when individual operations' execution is isolated, they suffer from additional delays, related to unknown resources, or effects similar to preemption-related delays.
We did not consider interference between the CPU and the \gpu, or other accelerators. Similar aspects, at least in terms of transparency and resilience, and methods should benefit to applications. 
The proposed requirements may be also lifted to reduce the impact of the method on the performance of applications. Operations from the same or different applications may notably be allowed to run concurrently provided they are known to not interfere with each other.
Applications or implementations tailored towards reduced execution time variability, e.g. relying on cooperative methods, should also be considered once a better understanding of the platform an applications has been established.
Finally, we need to consider the application of this work to support a platform-level resource manager such as \cite{rm:SCARLETT, actors, rm:DREAMS, rm:SECREDAS}.